Ferroelectric FET-based strong physical unclonable function: a low-power, high-reliable and reconfigurable solution for Internet-of-Things security

Xinrui Guo, Xiaoyang Ma, Franz Muller, Kai Ni, Thomas Kampfe, Yongpan Liu, Vijaykrishnan Narayanan, Xueqing Li

Abstract:

Hardware security has been a key concern in modern information technologies. Especially, as the number of Internet-of-Things (IoT) devices grows rapidly, to protect the device security with low-cost security primitives becomes essential, among which Physical Unclonable Function (PUF) is a widely-used solution. In this paper, we propose the first FeFET-based strong PUF exploiting the cycle-to-cycle (C2C) variation of FeFETs as the entropy source. Based on the experimental measurements, the proposed PUF shows satisfying performance including high uniformity, uniqueness, reconfigurability and reliability. To resist machine-learning attack, XOR structure was introduced, and simulations show that our proposed PUF has similar resistance to existing attack models with traditional arbiter PUFs. Furthermore, our design is shown to be power-efficient, and highly robust to write voltage, temperature and device size, which makes it a competitive security solution for Internet-of-Things edge devices.

**Introduction**

As the Internet-of-Things (IoT) technology develops, the demand for computing and storage on edge devices has increased substantially. Meanwhile, these edge devices are facing more security threats including hacking, information leaking, identity disguising, etc. [1], which makes it urgent to find solutions to not only software-level, but also hardware-level security protections. Considering the resource limitations of the edge devices, IoT security requires high density, high energy efficiency solutions. Among the various hardware security designs, physical unclonable function (PUF) is a promising hardware security primitive used for applications such as secret key generation and device authentication [2]-[4]. PUFs can map input challenges into device-unique random responses by utilizing the manufacturing process variations or device stochastic mechanisms. Fig. 1a. shows the conceptual authentication workflow using a PUF. The device-embedded PUF is produced by the manufacturer, while the challenge-response pairs (CRPs) are determined by the random entropy source rather than the manufacturer. Since the CRPs are random and unpredictable, they're used as the unique device fingerprint. Once the CRPs are determined, they need to be registered into the trusted cloud server. On the user's side, when authentication required, responses generated from some specific challenge inputs are compared with those pre-recorded ones on cloud to confirm the device identity. A good-to-have characteristic for PUF is the reconfigurability. In practical applications, sometimes the CRPs need to be reconfigured. For example, when a device changes its owner, then the fingerprints should be updated to avoid information leakage. Low-cost reconfigurability is a very competitive characteristic for IoT-PUF design.

Generally speaking, PUFs can be categorized into two types: strong PUF, whose CRP number is exponential to the PUF random unit number, and weak PUF, whose CRP number is quite limited, usually linear to the random unit number. Strong PUFs are more commonly used for authentication since they have enough CRPs, while weak PUFs are more likely to be used as a secret key generator.

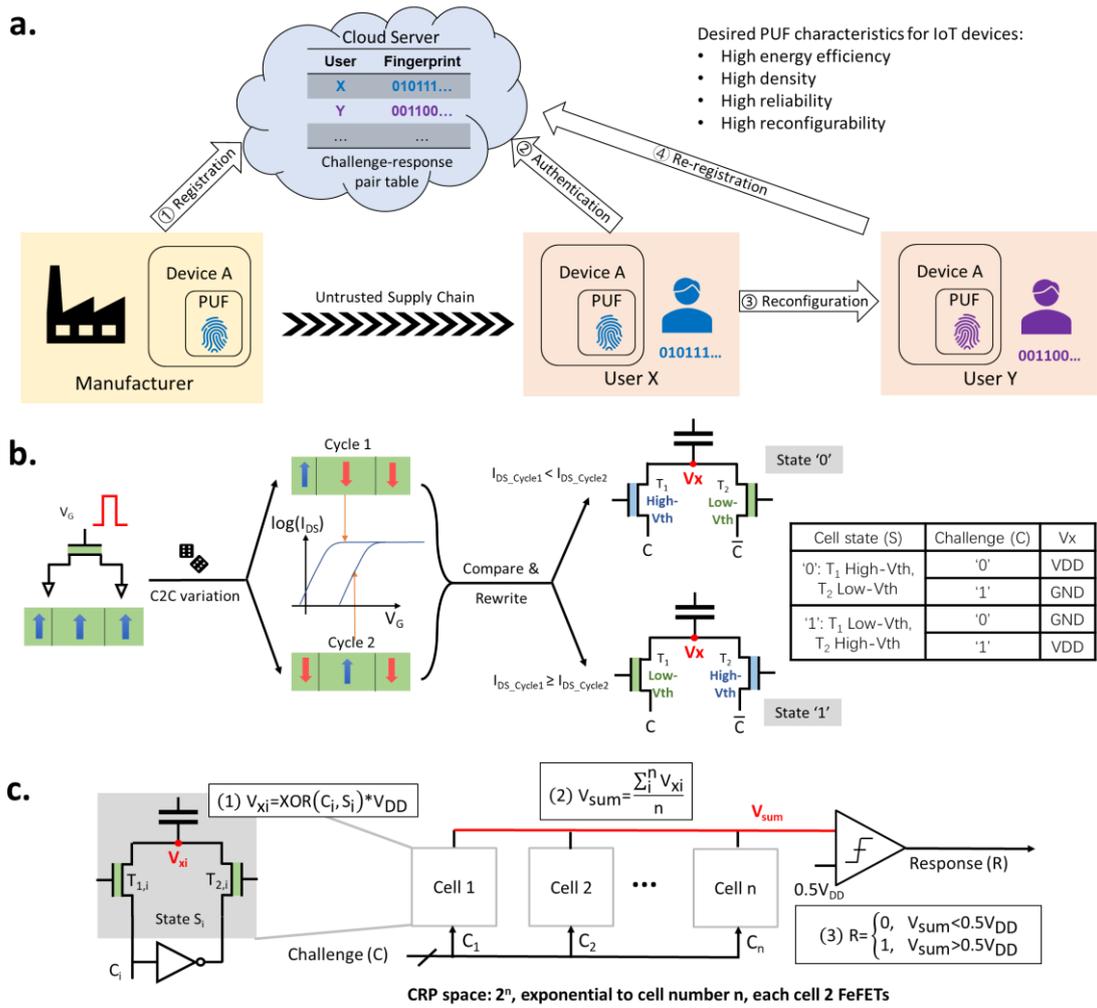

Fig. 1: FeFET strong PUF overview

a. Concept workflow of a PUF-based authentication scheme. The challenge-response pairs (CRPs) as fingerprints are registered on the trusted cloud server in advance. When authentication is required, the CRPs generated on the user's side are compared with the pre-recorded ones. If the CRPs need to be updated, the PUF can be reconfigured to have totally different CRPs, also the cloud database needs to be updated subsequently.

b. Entropy source and cell operations in the registration step of proposed FeFET strong PUF. Cycle-to-cycle (C2C) variations are exploited as the entropy source for our PUF design. During the registration, every cell with 2 FeFETs is rewritten to a certain state '0' or '1' according to the random polarization states difference of two write cycles. The Vx is determined by both cell state and input challenge bit.

c. Response generation of proposed FeFET strong PUF. The cells connected in a row

together generate one response bit. All Vxi's are coupled through the capacitors, equivalent to a voltage averaging operation to get Vsum. The response bit is finally generated by comparing Vsum with a reference voltage (0.5VDD).

In prior works, PUFs exploiting CMOS manufacturing process variations have been explored a lot, including path delay based arbiter PUF [5], oscillation frequency based ring-oscillator PUF [6], memory initial state based SRAM PUF [7], etc., which have good compatibility with other circuits. However, since the entropy sources these designs exploit are determined once the manufacturing has been completed, it's relatively harder to achieve high reconfigurability with low cost. Besides the above-mentioned traditional PUFs designs, non-volatile memories (NVM) provide a new chance for reconfigurable PUF design by exploiting device-internal stochastic mechanisms as the entropy source. Representative works include memristor based PUF [8], Phase Change Memory (PCM) based PUF [9], Spin-Transfer Torque Magnetic RAM (STT-MRAM) based PUF [10], carbon nanotube based PUF [11], Graphene Field-Effect Transistors (GFETs) based PUF [12] and so on. Most of these works utilize the device-to-device (D2D) and cycle-to-cycle (C2C) variations caused together by manufacturing process variations and device stochastic mechanisms, which gives them higher reconfigurability capability. However, most of these designs suffer from relatively low energy efficiency and high area overhead, and also need additional post-processing units to achieve high response reliability.

In recent years, Ferroelectric Field-Effect Transistors (FeFETs) has been an emerging non-volatile device which attracts much attention. In 2011, ferroelectricity was observed in doped-HfO2 materials [13], which brings interest in integrating it as a gate dielectric of FeFET with high CMOS compatibility. The non-volatility of the FeFET mainly comes from the ferroelectric layer, which can contain multiple ferroelectric domains [14]. These domains may stay at either positive polarization or negative polarization, causing the threshold voltage Vth change. The polarization states can be changed by applying a positive/negative voltage to the gate. Thanks to the voltage-mode writing operations, FeFETs have the advantage of ultra-low power consumption and high parallel capability compared with other common NVMs, suitable for IoT edge devices.

FeFETs are competitive candidates for hardware security design. The polarization switching of the ferroelectric domains have been observed to be a stochastic process [14]. This stochasticity together with the manufacturing process variations have bring obvious D2D and C2C variation, which is a very suitable entropy source for PUF design. What's more, ultra-low power writing operations make it easy to exploit C2C variation for PUF reconfigurability with little overhead. Prior works have tried to design FeFET-based weak PUFs [15], or optimize the traditional arbiter PUF by implementing the MUX with FeFETs to enhance reconfigurability [16], but as far as we know, no works on direct FeFET-based strong PUF design have been proposed yet.

In this article, we propose the first FeFET-based strong PUF design with high energy efficiency, high reliability and high reconfigurability. We exploit the obvious C2C variation of FeFET write operations as the PUF entropy source. Experimental measurements have confirmed that our proposed design shows excellent uniformity, uniqueness, reconfigurability and challenge length capability, as well as high robustness to write pulse amplitude, temperature, and device size. Simulations on machine learning attacks have also shown our proposed design have no worse ML attack performance than traditional arbiter PUFs, while high reconfigurability can further improve the attack resistance.

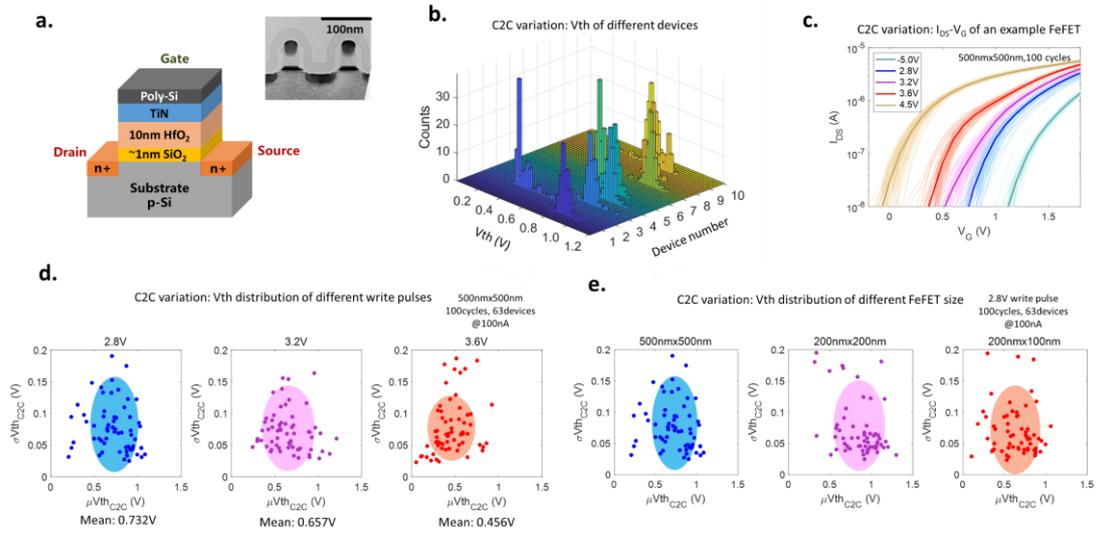

Fig. 2: FeFET device basics.

a. FeFET structure and TEM image

b-f. FeFET C2C variation.

b. Vth C2C distribution of different FeFET devices. The 500nmx500nm devices are erased by applying (-5V, 500ns) then written by applying (2.8V, 500ns) for a single write cycle. Vth extracted at 100nA with drain read voltage 100mV.

c. IDS-VG curves of different write cycles from a single example cell when applying 2.8/3.2/3.6V write pulse, and also the erased cells (-4.5V) and programmed cells (5.0V). As the write pulse amplitude rises, the curves go left, while the C2C variation stay obvious.

d. Vth C2C variation statistical results of 63 devices after different write pulses. Every dot in the figure represents a FeFET device. The shaded area represents the distribution mean and covers 75% of the points. As shown, for different devices the distribution is relatively scattered, but the mean value shows a falling trend when write pulse amplitude rises.

e. Vth C2C variation statistical results of different device sizes, each size 63 devices. The shaded area represents the distribution mean and covers 75% of the points. As the devices scale down, they still maintain significant C2C variation.

**FeFET basics: structure, characteristics and PUF design feasibility**

Ferroelectric materials have been used in memory for decades [22]. PZT-based ferroelectric capacitors were used in commercial ferroelectric RAM (FeRAM) as embedded NVM to replace eDRAM [17][18][19]. However, the FeRAM technology encountered issues of scaling down beyond 130nm with the PZT-based ferroelectric material system. Since the 2011 finding in [13], HZO-based materials have been found to be a good ferroelectric replacement with mature compatibility with the CMOS technology including the advanced FinFET technologies. Therefore, the ferroelectric layer could be integrated on the gate stack of the MOSFET, which is the concept of the aforementioned FeFET. After getting rid of the constraints by the PZT-based ferroelectric material system, the FeFET scalability has been demonstrated by recent reports on sub-10nm FeFETs [20][21].

The structure of FeFET is given in Fig. 2a, which could be considered as a conventional MOSFET with an extra ferroelectric layer sandwiched at the gate stack. The ferroelectric layer of FeFET consists of several ferroelectric domains, whose polarization states may stay at either positive or negative, and all these single domains determines the total polarization state. It has been observed that the polarization switching behavior of a single domain is an abrupt and stochastic process, regulated by the electric field across the domain [14]. This switching stochasticity together with the manufacturing process variations bring obvious D2D and C2C variation. This means that for different devices, they will have different $V_{th}$ even after applying the same write pulse; and for the same device, if we reset and then apply the same write pulse for several times, the $V_{th}$ also varies within some range. Fig. 2b gives the $V_{th}$ histogram of 10 different devices in 100 write cycles. As shown in the histogram, $V_{th}$ of different devices varies in a big range (0.2-1.2V), while $V_{th}$ of the same device distributes in a smaller range. This means that inter-device D2D variation is much larger than intra-device C2C variation. Note that the device-related bias plays a big role in $V_{th}$ distribution, which would cause the $V_{th}$ comparison result between devices not change over write cycles. If we exploit D2D variation as the entropy source by comparing different devices, the results may be stable, in other words, C2C variation will be hidden behind D2D variation, which seriously affect the reconfigurability. Taking this into account, we consistently exploit C2C variation as the PUF entropy source.

Fig. 2c shows the ID-VG curve families of a single FeFET, applying 2.8V/3.2V/3.6V-amplitude 500ns-wide write pulses after reset to high-Vth states. In the same plot, ID-VG curves of erased (applying -5.0V write pulse) and programmed (applying 4.5V write pulse) cells are also given. As seen, the curves with proper pulse amplitude all lie in the middle. The curves go left when the pulse amplitude rises, but C2C variation remains obvious. This can also show multiple level capability of FeFETs. Fig. 2d gives the Vth scatter plot of 63 devices in 100 write cycles. In these plots we can also see large D2D variation, as the mean Vth varies in a quite big range. But what's good for our PUF design is that, for every single device, C2C variation stays obvious under different write pulses. Similarly, Fig. 2e gives the C2C Vth scatter plot of devices with different sizes. We can confirm that even if the FeFET devices are scaled down, they still retain good C2C variation, which is quite exciting for scaling down design.

**FeFET strong PUF design: circuit and workflow**

As discussed above, we exploit the C2C variation of FeFET devices as the PUF entropy source. Fig. 1b conceptually shows how we do the PUF registration utilizing the C2C variation. The polarization states of the ferroelectric layer vary between different write cycles because of process variation and switching stochasticity, which brings randomness for PUF design. In our design, the PUF cell structure is quite simple, each cell consists of 2 FeFETs and 1 capacitor only. When repeated write cycles are applied to one FeFET in the cell, the unpredictable random polarization state difference between two write cycles caused by C2C variation can be exhibited by measurements, e.g., the read current IDS. After comparing the polarization states, the PUF cell is rewritten into a stable state according to the comparison: state '0' (T1 with high-Vth and T2 with low-Vth) or state '1' (T1 with low-Vth and T2 with high-Vth) to complete the registration. Note that the state of every cell will not change after a registration is completed, but the states are unpredictable before registration. The voltage Vx of internal node X used for response generation is determined together by cell state Si and the input challenge bit Ci, and the detailed correspondence is given in the table in Fig. 1b.

The response generation scheme is shown in Fig. 1c. One response bit is generated by N PUF cells in a row. Logically, the correspondence between Vxi, Ci and Si can be expressed as a XOR operation. All Vxi's are coupled through the capacitors with same capacitance, making an equivalent average operation to get Vsum. Finally, the Vsum is compared with a preset reference voltage to generate one-bit response. Note that all the $2^n$ n-bit challenges are available, implying the CRP space $2^n$ exponential to cell number n, and thus, our design is a strong PUF.

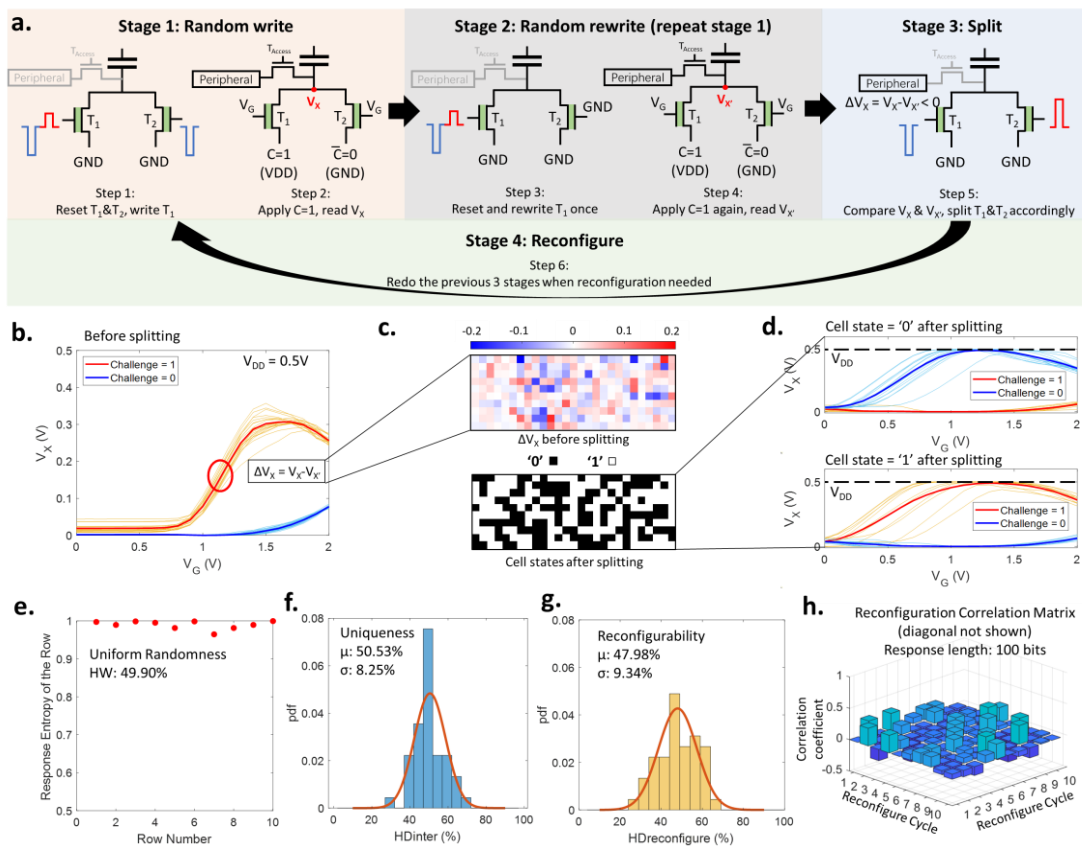

Fig. 3: FeFET strong PUF registration and performance:

a. Detailed cell registration workflow. Stages 1-3 make up the complete registration, while optional stage 4 repeats those 3 stages to complete the reconfiguration.

b-d. Visualized cells of an example registration.

b. VX-VG curves of 27 cells in 10 different write cycles, showing obvious C2C variation.

c. The calculated ΔVx map before rewrite (splitting) and the digital cell state pattern after rewrite (splitting).

d. VX-VG curves of the cells after splitting. The '1' and '0' cells give different curves, with the peaks very close to VDD, which is helpful for PUF reliability.

e. PUF uniformity, shown by calculating HW and the entropy of the response sequence (32-bit) generated by a row.

f. PUF uniqueness, shown by calculating HDinter.

g-h. PUF reconfigurability.

g. The calculated reconfigure HDinter. h. The 100-bit response correlation matrix of 10 reconfigure cycles, the ones on the diagonal are not shown for clarity.

A more detailed registration workflow is given in Fig. 3a. The whole workflow can be separated into 4 stages. Stage 1&2 are the same, which is to reset both T1 and T2 to high-Vth state then write T1 with a pre-defined pulse. Due to the C2C variation, T1 will change into some random polarization states in the two cycles. For the operation convenience, the polarization states are measured through Vx instead of IDS, and recorded in the peripheral circuit. After the first two stages, the Vx comparison is done in the peripheral, and two FeFETs are rewritten to high-Vth or low-Vth to form state '0' or '1', or in other words, split to enlarge the polarization difference. These 3 stages make up the complete registration workflow. Since all the above operations are done in voltage mode, our proposed design has relatively high energy efficiency compared with existing designs. In addition, the first two stages can be done simultaneously under high array-level parallelism due to the DC-power free FeFET write features.

What we want to highlight is that these stages are totally repeatable, which makes it possible to realize PUF reconfiguration easily with low cost. An extra stage 4 can be applied to repeat stages 1-3 for reconfiguration. Thanks to the unpredictable C2C variation, the Vx's in stages 1&2 would be different from the previous registration, and thus the final cell states are randomly set, making response bits almost irrelevant to those before the reconfiguration.

**FeFET strong PUF properties: basic PUF metrics and reconfigurability**

Experimental settings.

The electrical characterization of the PUF is performed on AND-connected FeFET arrays, consisting of 9 wordlines (WL) that combine the gates row-wise, and 7 bitlines/sourcelines (BL/SL), combining the drain and source contacts column-wise (see Supp. Fig 1). The control is carried out either by the use of Source-Measure-Units (SMUs) or Pin-Parametric-Measurement-Units (PPMUs), provided by a PXI-Express-System from National Instruments. Source selection happens via a custom switch matrix. Extended setup details, as well as general array-operation description and FeFET-preconditioning settings are given in [Methods]. A PUF-cell is constructed by shorting the SL of a left FeFET with the BL of a right FeFET by using externally connected jumper wires (see Supp. Fig 1). That way 3 PUF-elements can be constructed per WL, over the 9 WLs. The PUF characterization starts with the weak operation mode. Having the jumpers removed, the whole array is reset by a -5 V erase voltage, applied for 500 ns, resulting in all FeFETs being in the high Vt state. Afterwards all of the right sided FeFETs is weakly programmed, e.g. with a write voltage of 2.8V for 500 ns. To not disturb the erased high VT state of the left FeFET an inhibit-scheme is applied. Only the BL and SL of the actively to program FeFETs are tied to ground while inhibiting the BLs and SLs of the other FeFETs at 1.7 V. Furthermore the passive WLs are raised to 0.9 V to minimize any potential disturbs on the FeFETs on the passive WLs sharing the inhibited BLs and SLs. After writing the jumpers are connected. The PUF specific readout is performed by the PPMUs. The challenge C is set to be 0.5 V and C_bar equally 0 V. After applying the challenge a stepwise read is performed on the gates in a range from 0 V to 2 V in steps of 100 mV. While doing so the PPMU contacts connected to the shorted BL/SL contacts, representing the node Vx are in high ohmic mode and the node voltage Vx is measured. All 3 PUF-cells along the active WL are read in parallel. While reading all passive WLs are biased at -0.5 V to suppress leakage currents. C is first applied to the left FeFET. After reading C is applied to the right FeFET and the verification is repeated. To monitor influence of c2c-variations on the PUF-operation this routine is then repeated for a total of 10 times. The strong operation is performed the same way, but in addition after the weak write the result is processed in post-measurement verification. After resetting all FeFETs the condition with the lower Vx is strongly rewritten for full separation by applying the strong program voltage of 4.0 V for 500 ns. To protect the second FeFET from getting disturbed its BL and SL are inhibited to 2.6 V and the passive WLs are raised to 1.3 V.

Results and analysis

The PUF experimental results of 27 cells are given in Fig. 3. We have done 10 complete registrations for all these cells. The VX-VG curves in different write cycles of one example cell is given in Fig. 3b, with VDD set to 0.5V. At challenge bit = 1, where we extract Vx for later comparisons, obvious C2C variation can be observed. Fig. 3c shows the colormap of ΔVx, and the cell states after splitting of 27 cells x 10 registrations, which shows high randomness. The VX-VG curves of '0' cells and '1' cells after splitting are also given in Fig. 3d, showing large distance between the opposite states. A good thing for PUF reliability is that the peaks of Vx are very close to VDD, thanks to the FeFET high on-off ratio. This implies that after registration, the Vx's of all cells would be settled very close to GND or VDD with very low variance.

PUF basic metrics (more explanations see supplementary x) including uniform randomness and uniqueness are evaluated. To show PUF response uniform randomness, 100 random 27-bit challenges are chosen to generate a 100-bit response for 100 times, and the response entropy and Hamming Weight (HW) are calculated. As seen, the entropy are all close to ideal value 1, and the whole HW is calculated to be 49.90%, very close to ideal 50%. The inter-Hamming Distance (HDinter) of the above-mentioned responses are also calculated to confirm the uniqueness, and the results are given in Fig. 3f. The normalized HDinter are closely distributed around 50% ideal value, with 50.53% mean value and 8.25% standard deviation.

Experiments for reconfigurability analysis are also done, with results shown in Fig. 3g-h. The average normalized inter-Hamming Distance between responses in different reconfigurations generated from same challenge inputs are calculated to be 47.98% with 9.34% standard deviation, also closely distributed around 50% as shown in Fig. 3g. Correlation coefficient is also an important metric for reconfigurability. The correlation coefficients of the measured responses are drawn in Fig. 3h (the ones on the diagonal are hidden for clarity), all coefficients are close to 0, implying the responses are nearly independent, and the maximum correlation coefficient is under 0.3.

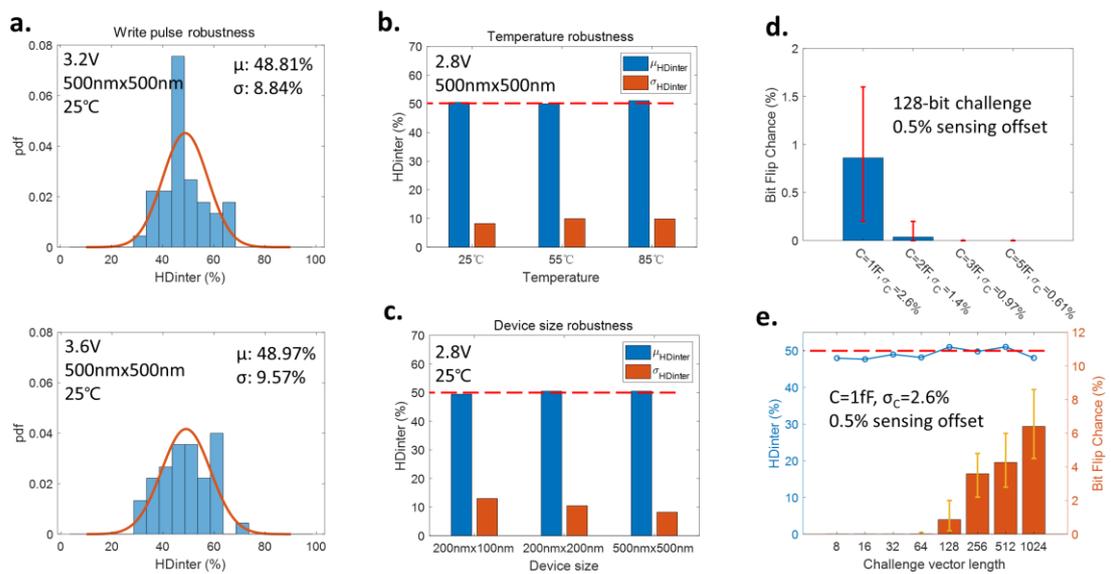

Fig. 4: FeFET strong PUF: parameter robustness and reliability.

a. Robustness to write pulse amplitude. 3 different write pulses (2.8V in Fig. 3, 3.2V and 3.6V) are applied to the same PUF array. And the calculated HDinter are given.

b. Temperature robustness. The experiments under 3 different temperatures (25, 55, 85℃) are done, with the calculated HDinter given.

c. Device size robustness. Same experiments are also done for 3 different FeFET device sizes (500nmx500nm, 200nmx200nm and 200nmx100nm), with the calculated HDinter given.

d-e. PUF reliability.

d. The 'Bit Flip Chance' caused by sensing offset under different capacitor mismatches. The σC are simulated from Monte Carlo simulations. Here 'bit flip' means the flipped bits compared with the theoretical 'ground truth' response. Since this flip is a static error, it won't cause any 'bit error' when generating response for several times.

e. The challenge vector length impact. As challenge goes longer, more static flips may happen, but they make no contribution to HDinter change.

**FeFET strong PUF: robustness and reliability**

In the practical applications for PUFs, there would be many non-idealities affecting PUF performance. On one hand, the parameters including pulse amplitude, device size and temperature would affect the FeFET operation characteristics, thus parameter robustness analysis is necessary to show wider scope of application for our design. On the other hand, response reliability (or sometimes called repeatability) is also an important metric to ensure we get exactly the same response every time we apply the same challenge input. Corresponding experiments are done to confirm the high robustness and high reliability of our proposed design, and the results are given in Fig. 4.

Fig. 4a shows the HDinter histogram with 3.2V/3.6V write pulse applied to the same PUF cells in Fig. 3, with 500nmx500nm device size and 25-degree temperature. As seen, the HDinter still lies around ideal 50% with similar small standard deviation. Similar analysis are also done for temperature (55 degree, 85 degree, shown in Fig. 4b) and device size (200nmx200nm, 200nmx100nm, shown in Fig. 4c), and they all give a satisfying HDinter, which proves that our proposed design is highly-robust to write pulse parameters, temperature and device size.

Response reliability is another essential metric to evaluate a PUF. An ideal PUF should generate stable responses for all available input challenges, in other words, the responses do not change over time. For many existing PUF designs, especially current-mode adding strong PUFs, they suffer from relatively high bit error rates. But for our proposed strong PUF, splitting strategy and voltage-mode adding ensure excellent response reliability. Though the capacitance mismatches and sensing amplifier offset may cause response generation error to the theoretical 'ground truth' response bit, the large cell state difference makes the average voltage Vsum have very little fluctuations, which makes no contribution to bit error in time dimension. Simulations on the capacitance mismatch and sensing offset effects have been done to check the design reliability, and the results are given in Fig. 4d. The capacitance mismatch data are simulated from Monte Carlo simulations, and the sensing offset is assumed to be 0.5% which is a common and reasonable value in practice. As the results show, smaller capacitors with larger mismatch may cause more response bits 'flip' from the theoretical 'ground truth' bit, but they are all static flipped bits and no bit errors are observed during repeat response generations, in other words the reliability is 100%. Another advantage of our design is the availability to have longer challenges. Since the $V_x$'s all lie close to GND or VDD, the final $V_{sum}$'s are distributed into several separated groups. As the challenge vector length n grows, the different $2^n$ $V_{sum}$ groups are distributed closer and closer. But if we can control the sensing offset at a low level which is not difficult for SA design, all $V_{sum}$'s can still be distinguished correctly. Fig. 4e gives the simulation results on different challenge vector length. As seen, when challenge goes longer, more static bit flips caused by sensing offset may happen, but they make no contribution to reliability degradation. Also, the calculated $HD_{inter}$ all stay close around 50%, implying stable excellent PUF performance. This is an exciting result which tells us we can make our challenge vector length grow with confidence. Longer challenge inputs are helpful to make higher PUF density and higher machine-learning attack resistance (discussed in the next section). Meanwhile, the proposed current-mode adding strong PUFs may suffer from power limitations and reliability degradations when making the challenge vector longer. Thanks to this, our design has very promising scalability, which means that higher PUF response density can be achieved without much performance loss.

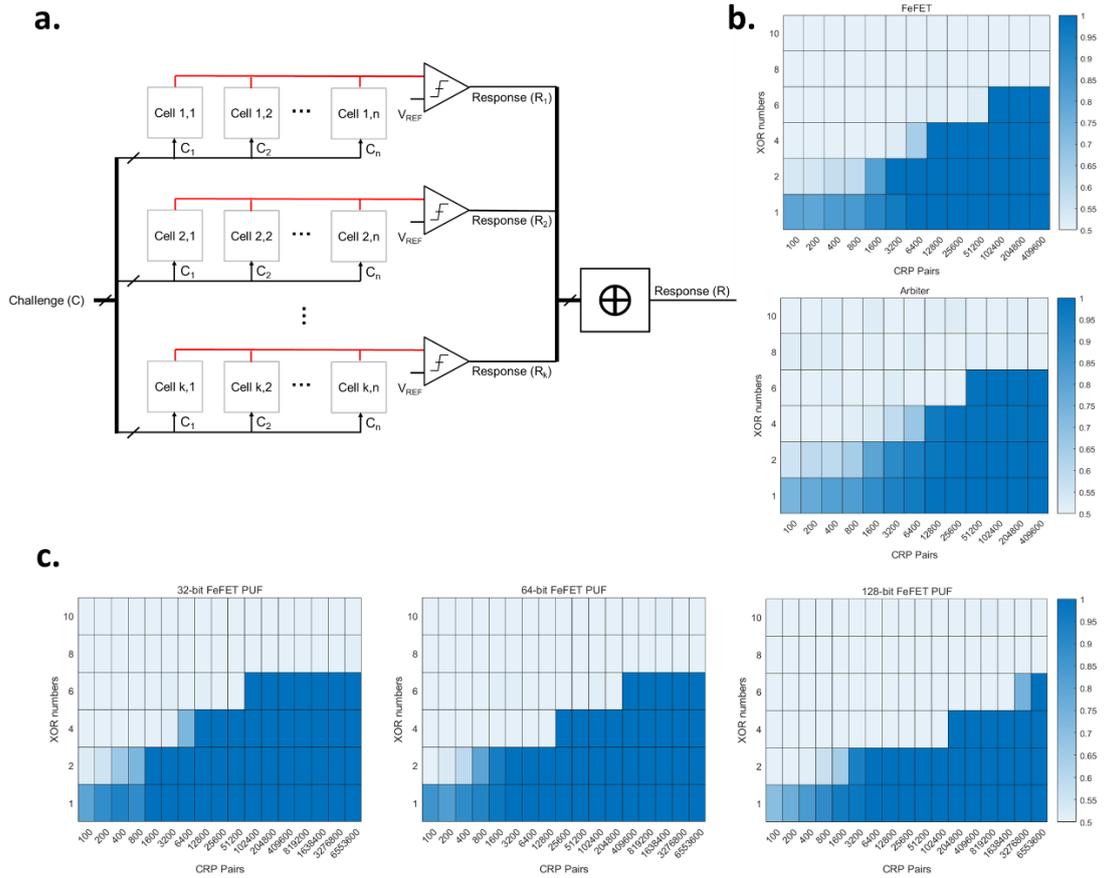

Fig. 5 Machine-learning attack resistant design

a. (n,k)-XOR structure used for machine-learning attack resistant design. k independent PUFs, each with n challenge bits are combined together with an XOR unit to introduce nonlinear complexity.

b. Prediction accuracy map using logic regression model (model parameters adjusted) of 27-bit-challenge proposed FeFET PUF and traditional arbiter PUF.

c. Prediction accuracy map of different challenge vector length.

**Machine-learning attack resistant design**

For PUFs, especially strong PUFs, machine learning attack resistance has been an essential metric. Typical machine learning attacks include logistic regression (LR) [23-26], support vector machines (SVM) [27], evolution strategies (ES) [23][28], deep learning (DL) [29]. A simple way to model both the arbiter PUF and our proposed PUF is by computing the scalar multiplication between the challenge vector and a PUF feature vector (each element in the feature vector represents the delay difference for arbiter PUF or cell capacitance for our proposed PUF, and the challenge vector contains ±1) [30]. The response bit is then determined by comparing the computed result with 0. Within this model, LR with high training efficiency is capable of achieving higher accuracy than SVM and ES strategies [23][24].

To improve machine learning resistance, XOR PUF [30][31], lightweight PUF [32][33], and interpose PUF [34] are introduced. Fig. 5a shows a common (n,k)-XOR structure containing k independent PUFs each with the same n-bit challenge input. The k output bits are then XORed together to produce the final output response bit. With a large k, XOR PUF could significantly increase the number of CRPs needed for training machine learning models with a density overhead and reliability compromise. As one of the most popular and effective method of improving machine learning resistance of PUF designs, we evaluate our proposed PUF design attacked by a simple LR algorithm. Our proposed design provides machine learning resistance as good as existing XOR arbiter PUFs based on our evaluation, and could limit the number of CRP pairs an attacker could acquire by reconfiguration.

We implement a LR model with Rprop as in [], where the feature vector is learned through a training CRP set. XOR arbiter PUF CRPs are generated by the pypuf toolbox [35], and the XORed proposed PUF CRPs are generated from the experimental data. The LR algorithm trains until the prediction accuracy converges on the training set, and the training process is repeated 10 times with different initial parameters in case the model is stuck at a local minimum. The model with the highest training set accuracy is chosen and tested on a test set with 10,000 CRP pairs. Fig. 5b shows the prediction accuracy on XOR arbiter PUF and our proposed PUF with different training set size and k (the number of independent PUFs XORed). With an increasing number of training CRP pairs, the LR model shows a transition from random guessing (accuracy near 0.5) to unstable guessing where the model sometimes falls into a local minimum, and finally stable prediction (accuracy near 1). The threshold training CRP pairs for these three stages are similar for XOR arbiter PUF and our PUF design, demonstrating similar machine learning resistance of the two designs.

Fig. 5c further shows the prediction results of the LR algorithm attacking PUFs with different challenge lengths. A longer challenge vector and larger XOR number could achieve in better machine learning resistance, requiring more CRP pairs for training, and the security of a reconfigurable PUF could be guaranteed by limiting the CRP pairs available to the attacker before a reconfiguration.

**Methods**

**FeFET Fabrication**

The AND-connected FeFET AND array structures are fabricated in GlobalFoundries' 28 nm high-k/metal gate technology node. The great advantage of the structures is the co-integrability with CMOS devices that has been demonstrated [Trentzsch, IEDM, 2016]. But the devices were also shown to be fabricable in GlobalFoundries' 22 nm fully-depleted silicon-on-insulator technology node [Dünkel, IEDM, 2017].

**FeFET characterization & setup**

The experimental characterization is performed on AND-connected FeFET arrays, consisting of 9 wordlines (WL), combining the gates row-wise, and 7 bitlines/sourcelines (BL/SL), combining the drain and source contacts column-wise. Array-control is enabled by a PXI-Express system from National Instruments. Each array-contact is selectively controllable by a NI PXIe-6570 Pin-Parametric Measurement-Unit (PPMU) and NI PXIe-4143 Source-Measure-Unit (SMU). The source can be selected by a custom switch-matrix for each contact. The switch-matrix itself is operated by additional PPMU-contacts. The matrix then connects to the array-structures via the probecard. Each FeFET is pre-conditioned for 50 program-erase-cycles with write-pulses of 4.5 V and -5 V at a pulse length of 500 ns each. Programming and erasing happens over the WL, utilized the SMUs, while keeping SLs and BLs at ground. For programming write voltages in the range of 2.8 V up to the full 4.5 V are utilized. Programming and erasing can be done fully parallel by a single pulse applied to all WLs at the same time as long as no selective inhibit conditions are required. This is the case for resets or the c2c-characterization. The FeFET read operation is done per WL with the SMUs by applying a voltage ramp from -0.2 V to 1.8 V in increments of 100 mV while measuring current at the drain terminals, which are biased at 100 mV. The bulk and source terminals are kept at ground. To ensure that there is sufficient time for charge detrapping before the first WLs VT states are read out and that the results remain comparable with the following WL readouts, a delay of 2 seconds is waited after each write pulse. The read operation takes approximately 1 ms per WL. If selective writing is necessary a VDD/3 oriented inhibit scheme [10.1109/TED.2013.2283465] is used. Where VDD corresponds to the WL-voltage, 2VDD/3 is applied to the FeFETs sharing the active WL and VDD/3 is applied to the passive WLs. To avoid disturbs it is ensured that inhibit voltages are applied from lowest to highest prior to applying the program voltage.

| Ref | This work | [1] | [2] | [3] | [4] | [5] | [6] |
|---|---|---|---|---|---|---|---|
| Technology | FeFET | GFET | RRAM | RRAM | CNT | STT-MRAM | CMOS Arbiter |
| Cell (core part) Structure | 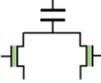 | 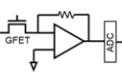 | 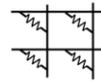 | 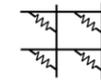 | 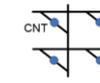 | 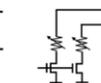 | 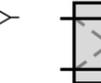 |
| Entropy Source | FeFET C2C variation | GFET carrier transport disorders | RRAM HRS variation | I-V nonlinearity variation | Geometry & placement | Geometry | Geometry |
| Weak or Strong | Strong | Weak | Weak & Strong | Strong | Weak | Weak | Strong |
| Uniformity (%) | 51.85 | ~50 | 49.02 | 49.5-50 | N/A | N/A | N/A |
| Uniqueness (%) | 50.53 | 47-50 | 48.3 | 50.0 | 50±0.39 | 50.0±0.1 | 47.13±0.44 |
| Reliability (%) | ~100 | 93-100 | 97.7-100 | ~97-98.9 | ~97 | ~100 | 96.96±0.08 |
| Reconfigurability (%) | 47.98 | 34-50 | 40.06 | N/A | N/A | N/A | N/A |
| Readout Latency | <1ns | ~100ns | N/A | 5ns | N/A | >10ns | N/A |
| Energy | ~3 fJ/bit | <12mW[a] | 0.102 pJ/bit | 20 fJ/bit | N/A | N/A | N/A |

a. Only power dissipation reported in the paper

Table I. Comparison with prior works.

References

[1] Zhao, K. & Ge, L. A survey on the internet of things security. In 2013 9th International Conference on Computational Intelligence and Security (CIS) 663–667, 2014.

[2] R. Pappu, B. Recht, J. Taylor, N. Gershenfeld, Physical one-way functions. Science 297, 2026–2030 (2002).

[3] C. Herder et al, "Physical Unclonable Functions and Applications: A Tutorial," in Proceedings of the IEEE, vol. 102, no. 8, pp. 1126-1141, 2014.

[4] Suh, G. E. & Devadas, S. Physical unclonable functions for device authentication and secret key generation. In Proc. 44th annual Design Automation Conference, 9–14, 2007.

[5] Daihyun Lim et al, "Extracting secret keys from integrated circuits," in IEEE Transactions on Very Large Scale Integration (VLSI) Systems, vol. 13, no. 10, pp. 1200-1205, Oct. 2005.

[6] A. Maiti et al, "A large scale characterization of RO-PUF," 2010 IEEE International Symposium on Hardware-Oriented Security and Trust (HOST), Anaheim, CA, 2010, pp. 94-99.

[7] A. Garg et al, "Design of SRAM PUF with improved uniformity and reliability utilizing device aging effect," 2014 IEEE International Symposium on Circuits and Systems (ISCAS), 2014, pp. 1941-1944.

[8]  Y. Pang et al., "25.2 A Reconfigurable RRAM physically unclonable function utilizing post-process randomness source with <6×10−6 native bit error rate," 2019 IEEE International Solid-State Circuits Conference - (ISSCC), San Francisco, CA, USA, 2019, pp. 402-404.

[9]  L. Zhang et al, "Exploiting process variations and programming sensitivity of phase change memory for reconfigurable physical unclonable functions," in IEEE Transactions on Information Forensics and Security, vol. 9, no. 6, pp. 921-932, June 2014.

[10] L. Zhang et al, "Highly Reliable Spin-Transfer Torque Magnetic RAM-Based Physical Unclonable Function With Multi-Response-Bits Per Cell," in IEEE Transactions on Information Forensics and Security, vol. 10, no. 8, pp. 1630-1642, Aug. 2015.

[11] Hu, Z., Comeras, J., Park, H. et al. Physically unclonable cryptographic primitives using self-assembled carbon nanotubes. Nature Nanotech 11, 559–565 (2016).

[12] Dodda, A., Subbulakshmi Radhakrishnan, S., Schranghamer, T.F. et al. Graphene-based physically unclonable functions that are reconfigurable and resilient to machine learning attacks. Nat Electron 4, 364–374 (2021).

[13] T. S. Böscke et al, "Ferroelectricity in hafnium oxide: CMOS compatible ferroelectric field effect transistors," 2011 IEDM, pp. 24.5.1-24.5.4.

[14] H. Mulaosmanovic et al., "Evidence of single domain switching in hafnium oxide based FeFETs: enabler for multi-level FeFET memory cells," 2015 IEEE International Electron Devices Meeting, 2015, pp. 26.8.1-26.8.3.

[15] X. Guo et al., "Exploiting FeFET Switching Stochasticity for Low-Power Reconfigurable Physical Unclonable Function," ESSCIRC 2021 - IEEE 47th European Solid State Circuits Conference (ESSCIRC), 2021, pp. 119-122.

[16] S. Ramanujam and W. Burleson, "Reconfiguring the Mux-Based Arbiter PUF using FeFETs," 2021 22nd International Symposium on Quality Electronic Design (ISQED), 2021, pp. 257-262.

[17] T. Mikolajick et al., "FeRAM technology for high density applications," Microelectronics Reliability, vol. 41, no. 7, pp. 947–950, 2001.

[18] H. Shiga, D. Takashima, et al., "A 1.6 gb/s ddr2 128 mb chain feram with scalable octal bitline and sensing schemes," IEEE Journal of Solid¬State Circuits, vol. 45, no. 1, pp. 142–152, 2010.

[19] M. Zwerg, A. Baumann, et al., "An 82μa/MHz microcontroller with embedded feram for energy¬harvesting applications," in 2011 IEEE International Solid¬-State Circuits Conference, 2011, pp. 334–336.

[20] A. Sharma and K. Roy, "1T non¬volatile memory design using sub¬10nm ferroelectric FETs," IEEE EDL, vol. 39, no. 3, pp. 359–362, 2018.

[21] Z. Krivokapic, U. Rana, et al., "14nm ferroelectric FinFET technology with steep subthreshold slope for ultra low power applications," in 2017 IEEE International Electron Devices Meeting (IEDM), IEEE, 2017, pp. 15–1.

[22] Khan, A.I., Keshavarzi, A. & Datta, S. The future of ferroelectric field-effect transistor technology. Nat Electron 3, 588–597 (2020).

[23] U. Rührmair, F. Sehnke, J. Sölter, G. Dror, S. Devadas, and J. Schmidhuber, "Modeling attacks on physical unclonable functions," in Proceedings of the 17th ACM conference on Computer and communications security, New York, NY, USA, Oct. 2010, pp. 237–249.

[24] G. T. Becker, "The Gap Between Promise and Reality: On the Insecurity of XOR Arbiter PUFs," in Cryptographic Hardware and Embedded Systems -- CHES 2015, Berlin, Heidelberg, 2015, pp. 535–555.

[25] Tobisch, Johannes, and Georg T. Becker. "On the scaling of machine learning attacks on PUFs with application to noise bifurcation." International Workshop on Radio Frequency Identification: Security and Privacy Issues. Springer, Cham, 2015.

[26] P. Santikellur and R. S. Chakraborty, "A Computationally Efficient Tensor Regression Network-Based Modeling Attack on XOR Arbiter PUF and Its Variants," IEEE Transactions on Computer-Aided Design of Integrated Circuits and Systems, vol. 40, no. 6, pp. 1197–1206, Jun. 2021.

[27] Daihyun Lim, J. W. Lee, B. Gassend, G. E. Suh, M. van Dijk and S. Devadas, "Extracting secret keys from integrated circuits," in IEEE Transactions on Very Large Scale Integration (VLSI) Systems, vol. 13, no. 10, pp. 1200-1205, Oct. 2005.

[28] Back, Thomas. Evolutionary algorithms in theory and practice: evolution strategies, evolutionary programming, genetic algorithms. Oxford university press, 1996.

[29] M. Khalafalla and C. Gebotys, "PUFs Deep Attacks: Enhanced modeling attacks using deep learning techniques to break the security of double arbiter PUFs," 2019 Design, Automation & Test in Europe Conference & Exhibition (DATE), 2019.

[30] J. Delvaux and I. Verbauwhede, "Side channel modeling attacks on 65nm arbiter PUFs exploiting CMOS device noise," 2013 IEEE International Symposium on Hardware-Oriented Security and Trust (HOST), 2013, pp. 137-142.

[30] G. E. Suh and S. Devadas, "Physical Unclonable Functions for Device Authentication and Secret Key Generation," 2007 44th ACM/IEEE Design Automation Conference, 2007, pp. 9-14.

[31] Chen Zhou, Keshab K. Parhi, and Chris H. Kim. "Secure and Reliable XOR Arbiter PUF Design: An Experimental Study based on 1 Trillion Challenge Response Pair Measurements, " In Proceedings of the 54th Annual Design Automation Conference (DAC) 2017.

[32] M. Majzoobi, F. Koushanfar and M. Potkonjak, "Lightweight secure PUFs," 2008 IEEE/ACM International Conference on Computer-Aided Design, 2008, pp. 670-673.

[33] E. Dubrova, O. Näslund, B. Degen, A. Gawell and Y. Yu, "CRC-PUF: A Machine Learning Attack Resistant Lightweight PUF Construction," 2019 IEEE European Symposium on Security and Privacy Workshops (EuroS&PW), 2019, pp. 264-271.

[34] Phuong Ha Nguyen, Durga Prasad Sahoo, Chenglu Jin, Kaleel Mahmood, Ulrich Rührmair and Marten van Dijk, "The interpose puf: Secure puf design against state-of-the-art machine learning attacks", IACR Transactions on Cryptographic Hardware and Embedded Systems, pp. 243-290, 2019.

[35] Nils Wisiol et al. "pypuf: Cryptanalysis of Physically Unclonable Functions (Version 2)", Zenodo, June 2021.